\documentclass[lettersize,journal]{IEEEtran}
\usepackage{amsmath,amsfonts}
\usepackage{algorithmic}
\usepackage{algorithm}
\usepackage{array}
\usepackage[caption=false,font=normalsize,labelfont=sf,textfont=sf]{subfig}
\usepackage{textcomp}
\usepackage{stfloats}
\usepackage{url}
\usepackage{verbatim}
\usepackage{graphicx}
\usepackage{cite}
\begin{document}

\title{Design and Performance of 220 and 270 GHz\\ Bandpass Filters for BICEP Array
}
\author{A.~Steiger$^{a*}$,
BICEP/\textit{Keck}~Collaboration,
P.~A.~R.~Ade$^{b}$,
Z.~Ahmed$^{c,d}$,
M.~Amiri$^{e}$,
D.~Barkats$^{f}$,
R.~Basu~Thakur$^{a}$,
C.~A.~Bischoff$^{g}$,
D.~Beck$^{h}$,
J.~J.~Bock$^{a,i}$,
H.~Boenish$^{f}$,
V.~Buza$^{j}$,
K.~Carter$^{f}$,
J.~R.~Cheshire~IV$^{k,a}$,
J.~Connors$^{l}$,
J.~Cornelison$^{f}$,
L.~Corrigan$^{f}$,
M.~Crumrine$^{m}$,
S.~Crystian$^{f}$,
A.~J.~Cukierman$^{h}$,
E.~Denison$^{l}$,
L.~Duband$^{n}$,
M.~Echter$^{f}$,
M.~Eiben$^{f}$,
B.~D.~Elwood$^{o,f}$,
S.~Fatigoni$^{a}$,
J.~P.~Filippini$^{p,q}$,
A.~Fortes$^{h}$,
M.~Gao$^{a}$,
C.~Giannakopoulos$^{g}$,
N.~Goeckner-Wald$^{h}$,
D.~C.~Goldfinger$^{h}$,
J.~A.~Grayson$^{h}$,
A.~Greathouse$^{a}$,
P.~K.~Grimes$^{f}$,
G.~Hall$^{m,h}$,
G.~Halal$^{h}$,
M.~Halpern$^{e}$,
E.~Hand$^{g}$,
S.~A.~Harrison$^{f}$,
S.~Henderson$^{c,d}$,
J.~Hubmayr$^{l}$,
H.~Hui$^{a}$,
K.~D.~Irwin$^{h}$,
J.~H.~Kang$^{a}$,
K.~S.~Karkare$^{c,d}$,
S.~Kefeli$^{a}$,
J.~M.~Kovac$^{o,f}$,
C.~Kuo$^{h}$,
K.~Lau$^{a}$,
M.~Lautzenhiser$^{g}$,
A.~Lennox$^{q}$,
T.~Liu$^{h}$,
K.~G.~Megerian$^{i}$,
M.~Miller$^{f}$,
L.~Minutolo$^{a}$,
L.~Moncelsi$^{a}$,
Y.~Nakato$^{h}$,
H.~T.~Nguyen$^{i,a}$,
R.~O'Brient$^{i,a}$,
S.~Paine$^{f}$,
A.~Patel$^{a}$,
M.~A.~Petroff$^{f}$,
A.~R.~Polish$^{o,f}$,
T.~Prouve$^{n}$,
C.~Pryke$^{m}$,
C.~D.~Reintsema$^{l}$,
T.~Romand$^{a}$,
D.~Santalucia$^{f}$,
A.~Schillaci$^{a}$,
B.~Schmitt$^{f}$,
E.~Sheffield$^{f}$,
B.~Singari$^{m}$,
K.~Sjoberg$^{f}$,
A.~Soliman$^{i,a}$,
T.~St~Germaine$^{f}$,
B.~Steinbach$^{a}$,
R.~Sudiwala$^{b}$,
K.~L.~Thompson$^{h,c}$,
C.~Tsai$^{f}$,
C.~Tucker$^{b}$,
A.~D.~Turner$^{i}$,
C.~Vergès$^{f}$,
A.~G.~Vieregg$^{j}$,
A.~Wandui$^{a}$,
A.~C.~Weber$^{i}$,
J.~Willmert$^{m}$,
W.~L.~K.~Wu$^{d,c}$,
H.~Yang$^{h}$,
C.~Yu$^{j}$,
L.~Zeng$^{f}$,
C.~Zhang$^{c}$,
S.~Zhang$^{a}$
\\ 
\vspace{5mm}
$^{a}$ Department of Physics, California Institute of Technology, Pasadena, CA 91125, USA\\
$^{b}$ School of Physics and Astronomy, Cardiff University, Cardiff, CF24 3AA, United Kingdom\\
$^{c}$ Kavli Institute for Particle Astrophysics and Cosmology, Stanford University, Stanford, CA 94305, USA\\
$^{d}$ SLAC National Accelerator Laboratory, Menlo Park, CA 94025, USA\\
$^{e}$ Department of Physics and Astronomy, University of British Columbia, Vancouver, British Columbia, V6T 1Z1, Canada\\
$^{f}$ Center for Astrophysics, Harvard \& Smithsonian, Cambridge, MA 02138, USA\\
$^{g}$ Department of Physics, University of Cincinnati, Cincinnati, OH 45221, USA\\
$^{h}$ Department of Physics, Stanford University, Stanford, CA 94305, USA\\
$^{i}$ Jet Propulsion Laboratory, California Institute of Technology, Pasadena, CA 91109, USA\\
$^{j}$ Kavli Institute for Cosmological Physics, University of Chicago, Chicago, IL 60637, USA\\
$^{k}$ Minnesota Institute for Astrophysics, University of Minnesota, Minneapolis, MN 55455, USA\\
$^{l}$ National Institute of Standards and Technology, Boulder, CO 80305, USA\\
$^{m}$ School of Physics and Astronomy, University of Minnesota, Minneapolis, MN 55455, USA\\
$^{n}$ Service des Basses Températures, Commissariat à l'Énergie Atomique, 38054 Grenoble, France\\
$^{o}$ Department of Physics, Harvard University, Cambridge, MA 02138, USA\\
$^{p}$ Department of Physics, University of Illinois at Urbana--Champaign, Urbana, IL 61801, USA\\
$^{q}$ Department of Astronomy, University of Illinois at Urbana--Champaign, Urbana, IL 61801, USA
}
\markboth{}%
{Shell \MakeLowercase{\textit{et al.}}: A Sample Article Using IEEEtran.cls for IEEE Journals}


\maketitle

\begin{abstract}
The BICEP Array (BA) is the latest in the BICEP/\textit{Keck} series of experiments that aim to measure the polarization of the cosmic microwave background (CMB) with small aperture polarimeters located at the South Pole. To constrain the frequency response of these receivers, each detector is serially coupled to a band-pass filter (BPF). The electric circuits of these BPFs utilize series and shunt capacitors as well as series inductors, but critically do not include shunt inductors which simplifies fabrication. The filters are designed and simulated with \textit{Sonnet}, and optimized for noise by considering loading from the atmosphere and the CMB. Multiple 220 GHz detector modules have had their frequency response measured at the South Pole. The 270 GHz detector modules have recently begun testing in a lab setting, and their performance in a BA receiver will be measured this winter.
\end{abstract}


\section{Introduction}
\IEEEpubidadjcol
The BICEP Array (BA) is the latest in the BICEP/\textit{Keck} series of experiments that aim to measure the polarization of the cosmic microwave background (CMB) with small aperture polarimeters located at the South Pole\cite{BK18}. These polarization measurements will allow for new constraints on the physics of cosmic inflation\cite{CMBS4,Seljak1997,Kamionkowski1996,Seljak1997_2}. BA is a collection of BICEP3-style receivers\cite{Hui2019,Chesire2024,Nakato2024}. As of September 2025, there are two completed BA receivers: one observing at 30/40 GHz and one at 150 GHz, and there is one partially completed receiver observing at 220/270 GHz. The 150 GHz receiver is most sensitive to the CMB, while the 30/40 GHz and 220/270 GHz receivers are useful for constraining foregrounds\cite{foregrounds}. The 30/40 GHz receiver constrains synchrotron radiation and the 220/270 GHz receiver constrains polarized dust.

The BA detectors are transition edge sensors (TES), which utilize the steep temperature versus resistance property of superconductors to function as highly sensitive bolometers. A detector module consists of 648 TESs and a full receiver focal plane houses 12 modules, resulting in tens of thousands of detectors across the full experiment. Each TES is serially coupled to a band-pass filter (BPF) to define the frequency response of the detectors. This work details the design, optimization, and performance of these BPFs.

\section{Design}
The BPFs used in the 220 and 270 GHz BA modules are based on the work of Galbraith and Rebeiz\cite{Cochlea}, which details how a standard passband filter can be transformed into a series of pi networks of capacitors and inductors. The design process begins with a set of ideal circuit element values for an equal-ripple 3-pole low-pass filter with 0.5 dB of ripple. We use values found in table 8.4 of \textit{Microwave Engineering} by Pozar \cite{Pozar} and then convert them into values for a BPF as described in chapter 8 of the same text. We then transform the circuit into a pi-network circuit, and tune the response by tuning the circuit element values. This pi-network design utilizes series and shunt capacitors as well as series inductors, but no shunt inductors. This is one of the main motivations of this design as shunt inductors are more difficult to consistently fabricate. This style of filter has been shown to work in SPT-3G and Simons Observatory\cite{SPT3G,SO}. 

This filter design is electrically similar to the T-network designs used in BICEP3, BICEP2, and \textit{Keck} Array \cite{BICEP3,BICEP2}. The T-network design is also realized without shunt inductors, the main advantage of the pi-network over the T-network is that we have found it easier to fabricate the optimal values for the circuit elements with the pi-network. 

Another popular filter option is the resonant stub filter, which uses quarter-wave sections of transmission lines, such as in ACTpol\cite{ACTpol}. The primary reason we prefer the pi-network design is that in this design the inductance and capacitance
values are primarily determined by lithographic geometry, more so than in the quarter-wave stubs which are sensitive to impurities in the film\cite{RogerThesis}. 

\begin{figure}[!h]
\centering
\includegraphics[width=3in]{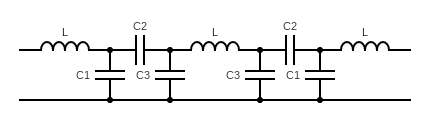}
\caption{A circuit diagram of the pi-network BPF design. L is a series inductor, C1 is the outer shunt capacitor, C2 is the series capacitor, and C3 is the inner shunt capacitor.}
\label{circuit_diagram}
\end{figure}

A circuit diagram of the pi-network filter design is shown in Fig. \ref{circuit_diagram}, it is a third order-filter.

The filter is made with Nb microstrip, separated from a Nb ground plane by a layer of $\textrm{SiO}_2$. The inductors (L in Fig. \ref{circuit_diagram}) are realized as short sections of high impedance line with a cut out in the ground plane beneath them. The structure is essentially a co-planar waveguide with the central line offset from the ground plane. Such a transmission line will have an inductance given by
\begin{align}
    L=\frac{Zl}{v}  
\end{align}
Here $Z$ is the impedance, $l$ is the length of the line, and $v$ is the wave speed in the microstrip line. The key here is that the inductance can be tuned by making the line shorter or longer.

The shunt capacitors (C1 and C3 in Fig. \ref{circuit_diagram}) are parallel plate capacitors with the ground plane serving as one of the plates. There are two shunt capacitor types used in the design, one inner and one outer as shown in Fig. \ref{circuit_diagram}. The capacitance is a function of the area of the plate, so we have the length and width to use as tuners.

The series capacitors (C2 in Fig. \ref{circuit_diagram}) are also parallel plate, with a carved-out section of the ground plane serving as one of the plates. This is done to keep fabrication simple as it allows us to create the whole filter with two Nb films. The gap in the ground plane around the lower plate can produce parasitic inductance, so we keep it as small as we can reliably fabricate. In practice, we can also compensate by lowering the inductance of the series inductors. The upper plate is split in half to minimize RF vias.

\begin{figure}[!h]
\centering
\includegraphics[width=3.3in]{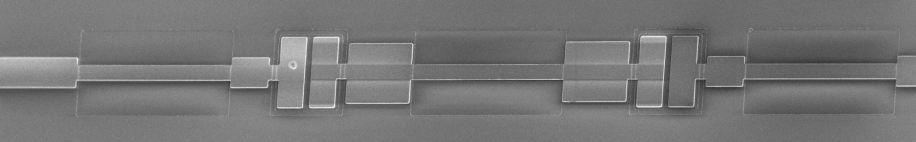}
\caption{An SEM image a of fabricated 220 GHz filter.}
\label{real_filter}
\end{figure}

The filters are designed and simulated with \textit{Sonnet}, the simulation geometries are shown below in Fig. \ref{sonnet_design} and a scanning electron microscope (SEM) image of a fabricated 220 GHz filter is shown in Fig. \ref{real_filter}.

\begin{figure}[!h]
\centering
\includegraphics[scale=0.295]{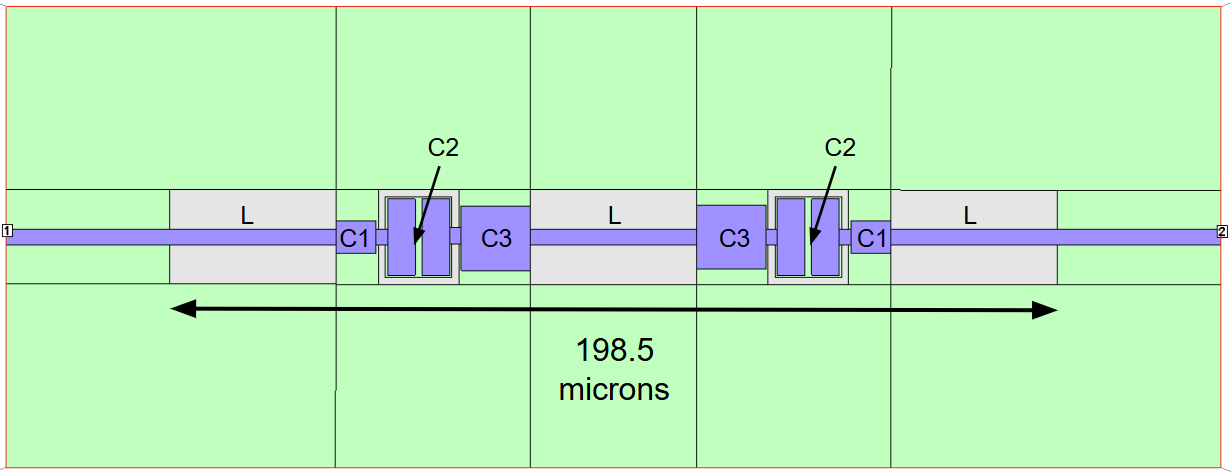}
\includegraphics[scale=0.295]{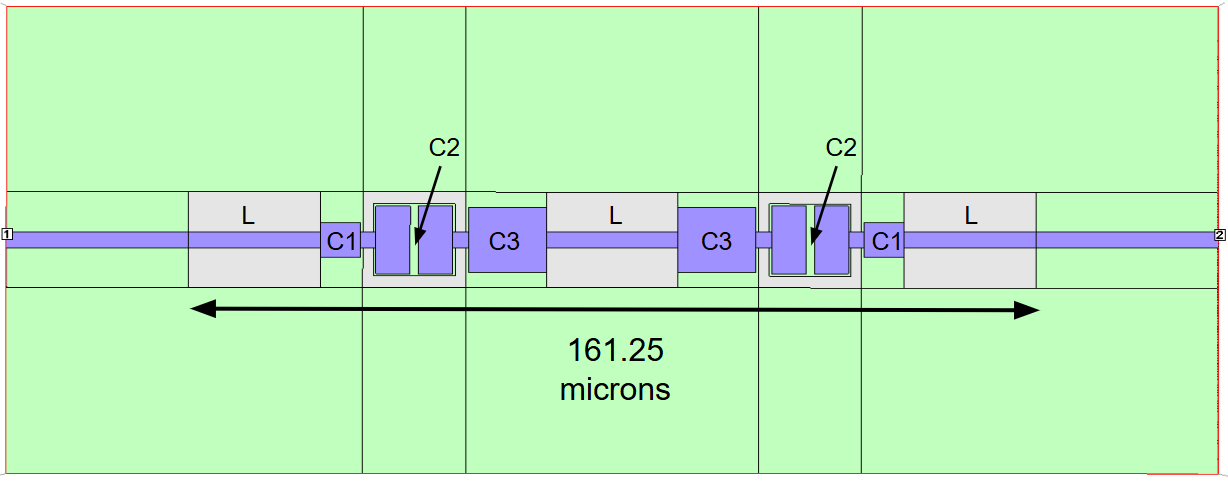}
\caption{The 220 GHz (top) and 270 GHz (bottom) filter designs in \textit{Sonnet} for simulation. The green is the ground plane, and the blue is the microstrip layer. The gray area is where the ground plane is cut out. The circuit elements are labeled as in Fig. \ref{circuit_diagram}, and the lengths of the filters are also denoted.}
\label{sonnet_design}
\end{figure}

\begin{figure}[]
\centering
\includegraphics[scale=0.47]{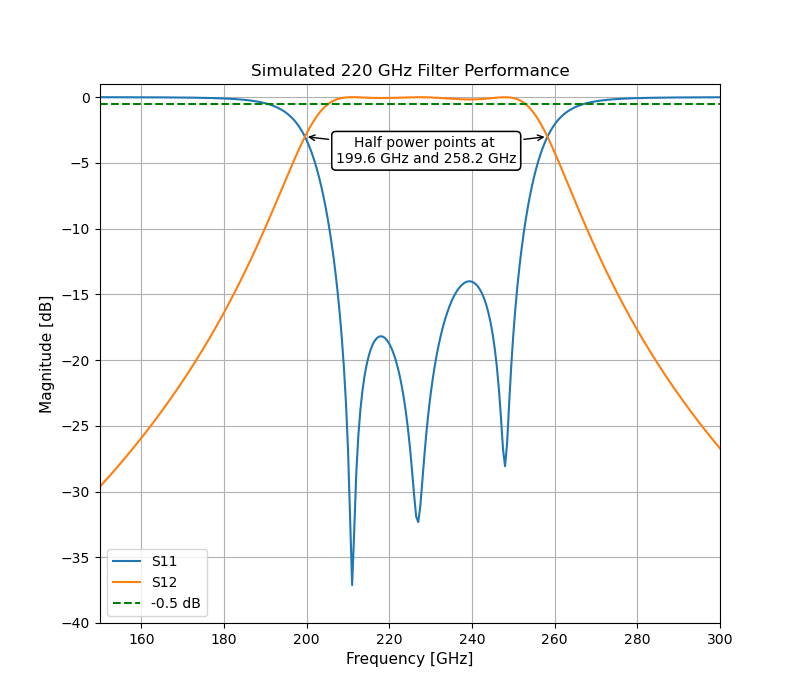}
\includegraphics[scale=0.47]{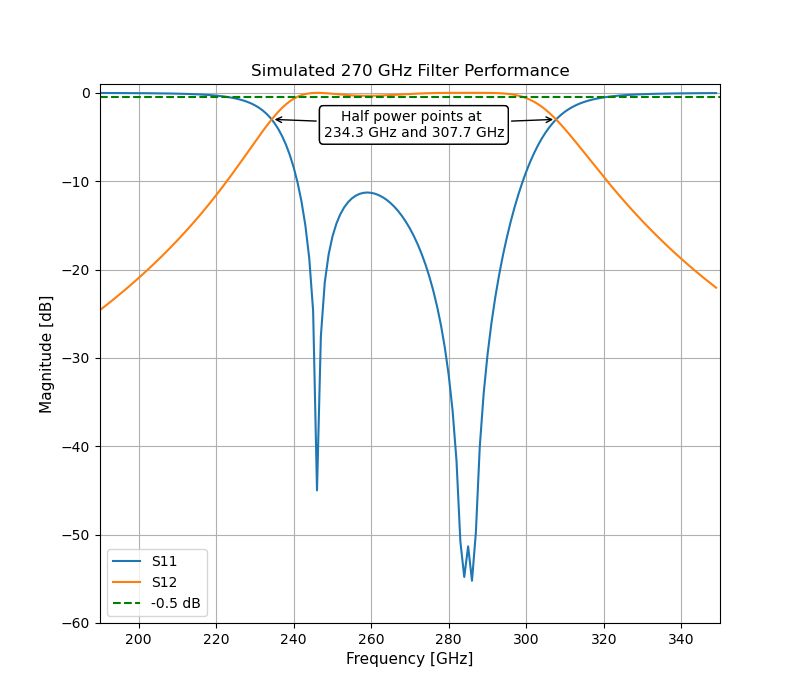}
\caption{The simulated S11 and S12 curves for the 220 GHz (top) and 270 GHz (bottom) filters. The in band ripple does not exceed 0.5 dB as denoted by the dashed green line. The half power points at -3 dB are also noted.}
\label{sonnet_performance}
\end{figure}

The dielectric layer stack used for \textit{Sonnet} simulations from the bottom up is 400 microns of silicon, 1 micron of silicon nitride, 0.3 microns of $\textrm{SiO}_2$, and then 400 microns of vacuum on top.  The relative permittivity values used are 11.9 for the silicon, 7.5 for the silicon nitride, and 4.0 for the $\textrm{SiO}_2$, and all the dielectrics are assumed to have no loss. The Niobium ground plane sits on top of the silicon nitride, and the microstrip layer sits above the $\textrm{SiO}_2$. The niobium is assumed to be a lossless conductor, and it is given a surface inductance of 0.1 pH/sq to simulate kinetic inductance. The simulated S11 and S12 of the filters along with the half power points are shown in Fig. \ref{sonnet_performance}.

The 3-pole response is apparent for the 220 GHz filter, for the 270 GHz filter the two higher frequency poles are overlapping. Both filters achieve reflection (S11) less than -10 dB across the band. Compared to the initial 0.5 dB equal-ripple circuit parameters, we note that neither the circuit transformation nor filter optimizations pushed the ripple in the passband (S12) beyond 0.5 dB, with the maximum deviation from 0 db in the passband being 0.18 for the 220 GHz filter and 0.34 for the 270 GHz filter. The motivation for the chosen band properties are discussed in the following section.
\section{Optimization}
\label{Optimization}
The filters are noise optimized by considering the noise equivalent temperature (NET) from the loading of a model atmosphere and the CMB, as a function of the filter response. The NET is calculated as follows, we start by defining the optical power from both sources: 
\begin{align}
\label{Q_cmb}
\mathrm{Q_{cmb}} &= \int{d\nu}\ \eta_{\mathrm{telescope}}\left(\frac{h\nu}{e^{x_{\mathrm{cmb}}}-1}\right)(1-\eta_{\mathrm{atm})}S(\nu) \\
\label{Q_atm}
\mathrm{Q_{atm}} &= \int{d\nu}\ \eta_{\mathrm{telescope}}\left(\frac{h\nu}{e^{x_{\mathrm{atm}}}-1}\right)S(\nu)
\end{align}
with 
\begin{align}
x_{\mathrm{cmb}}\equiv\frac{h\nu}{k_BT_{\mathrm{cmb}}},x_{\mathrm{atm}}\equiv
\frac{h\nu}{k_BT_{\mathrm{sky}}},T_{\mathrm{sky}}=\eta_{\mathrm{atm}}(\nu)T_{\mathrm{atm}} \nonumber
\end{align}
where $T_{\mathrm{sky}}$ is the brightness temperature in Kelvin, $\eta_{atm}(\nu)$ is the atmospheric efficiency (or $1-\textrm{atmospheric transmission)}$, and $S(\nu)$ is the filter power efficiency (analogous to $|S21|^2$), and taking $T_{\mathrm{cmb}}=2.7\mathrm{K},\ T_{\mathrm{atm}}=250\mathrm{K},\ \eta_{\mathrm{telescope}}=0.5$. The integrals in equations \ref{Q_cmb} and \ref{Q_atm} run over all frequencies but are negligible outside the relevant frequencies governed by the source temperatures $T_{\mathrm{cmb}}$ and $T_{\mathrm{atm}}$ and by  $S(\nu)$. From equation \ref{Q_atm} onwards, we assume the Rayleigh-Jeans limit.
Then we calculate noise equivalent power(NEP) as
\begin{align}
    \mathrm{NEP}^2=2h\langle\nu\rangle{Q_\mathrm{tot}}+\frac{2{Q_\mathrm{tot}}^2}{\Delta\nu}
\end{align}
with
\begin{align}
\label{NET_defs}
    \langle\nu\rangle \equiv \frac{\int{S(\nu)\nu{d}\nu}}{\int{S(\nu){d}\nu}}, {\Delta\nu}\equiv\frac{\left(\int{S(\nu){d}\nu}\right)^2}{\int{S(\nu)^2{d}\nu}}
\end{align}
and $Q_\mathrm{tot}$ being the sum of $Q_\mathrm{cmb}$ and $Q_\mathrm{atm}$. Then we convert to NET as
\begin{align}
\label{NET}
    \mathrm{NET}=\mathrm{NEP}\left(\frac{dQ_\mathrm{CMB}}{dT}\right)^{-1}
\end{align}
Here looking at $Q_\mathrm{CMB}$ and not $Q_\mathrm{tot}$ for the conversion derivative so that we can express NET in thermodynamic CMB units. 

The atmosphere model used for the NET calculations represents a Mauna Kea atmosphere with a PWV of 0.30mm and an altitude of 4,200 meters. The PWV level is comparable to the South Pole and while the altitude is higher, this will not change the spectral locations of atmospheric emission lines which are the critical feature for optimizing the filter pass bands\footnote{The lower pressure at Mauna Kea does result in slightly smaller atmospheric emission lines due to pressure broadening, but this effect is negligible in optimizing the filter band.}. In line with our filter design, the filter model used is an ideal 0.5 dB equal-ripple 3-pole Chebyshev filter. 

The noise minimum is found through a simple grid search over band center and bandwidth. In this section, the band center is defined as halfway between the two half power points, and bandwidth is the distance between the half power points. This simple definition is suitable for the optimization process as specifying a band center and bandwidth for a given filter model completely defines the filter.

For 220 GHz, we try to get as close to the noise minimum as possible. There is a limited precision with which we can tune the band center and bandwidth as functions of the capacitors and inductors. We arrive at a band center of 229 GHz and a fractional bandwidth of 0.256 which achieves an NET of 187.9 $\mu{K}_{\mathrm{cmb}}/\sqrt{\mathrm{Hz}}$ which is within 1\% of the grid search minimum of 188.5.

\begin{figure}[]
\centering
\includegraphics[scale=0.53]{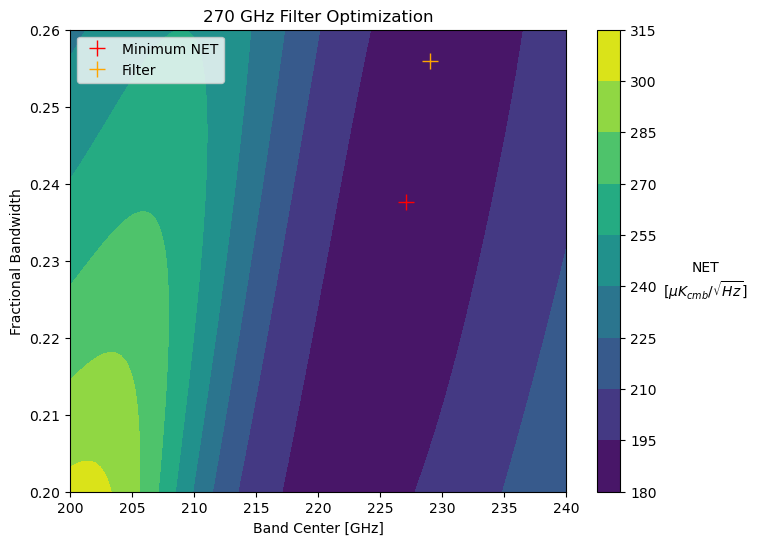}
\includegraphics[scale=0.53]{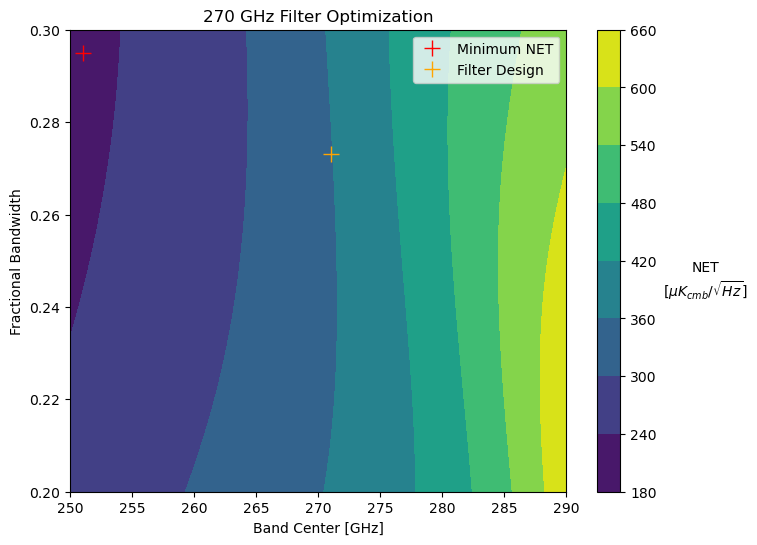}
\caption{The 220GHz (top) and 270GHz (bottom) NET color plots for filter optimization.}
\label{NET color}
\end{figure}

For 270 GHz, equations \ref{Q_cmb} and \ref{Q_atm} push the band center and bandwidth to the low and high ends of the grid search, respectively, because the atmosphere is less transparent around this frequency compared to 220 GHz. We prefer to keep the band center around 270 GHz as the free parameters of the dust model used in the mainline BICEP analysis likelihood search are better constrained with multiple frequency measurements \cite{BK18}. It is also difficult in practice to fabricate this type of filter to have a fractional bandwidth much higher than 0.27. We arrive at a band center of 271 GHz and a fractional bandwidth of 0.273, which has an NET of 358.3 $\mu{K}_{\mathrm{cmb}}/\sqrt{\mathrm{Hz}}$. This is ~1.6 times higher than the grid search minimum of 223.8 but it is more in line with our science goals for this receiver. Also, the photon noise within the band defined by the filters is not the dominant source of noise in the receiver, so a higher NET filter does not result in a significant penalty for our science results. NET color plots showing the filter specifications and the NET minimum are shown in Fig. \ref{NET color}.

\section{Performance}
\begin{figure}[]
\centering
\includegraphics[scale=0.6]{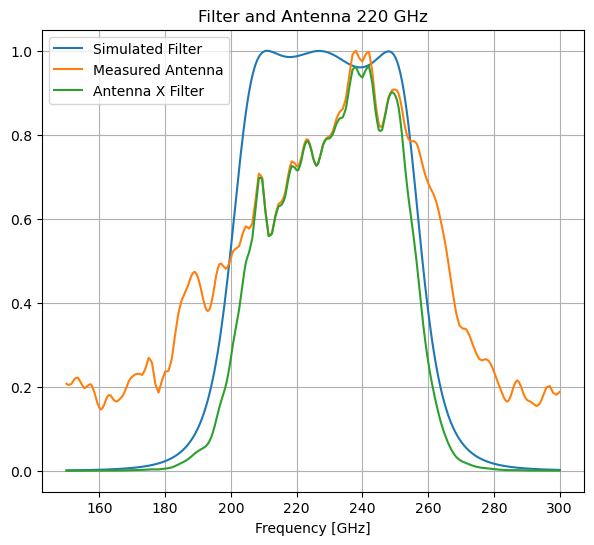}
\caption{The simulated BPF $|\textrm{S}21|^2$ for 220 GHz (blue), antenna spectrum from loss test detectors (orange), and antenna and filter product (green).}
\label{filterxantenna}
\end{figure}

Three 220 GHz modules were installed at the South Pole during the 2024-25 winter. Their spectral performance was measured in-situ with a Martin–Puplett Fourier Transform Spectrometer (FTS)\cite{MP_FTS}. We should note that the spectra obtained from FTS measurements are not direct measurements of the BPFs because the antenna also has a non-flat frequency response. BA modules use dual-slot polarization antennas which have a proven history of success for BICEP/\textit{Keck} experiments\cite{Antenna,BICEP3}. We can measure the passband of the antennas by looking at the FTS results of loss test detectors. Each BA detector module has two pairs of loss test detectors, which are detectors without a BPF. Their main purpose is to measure loss in the Nb microstrip and for that it is preferable to look at loss across all frequencies that the antenna is sensitive to. The loss measurements are beyond the scope of this work, but the lack of a BPF means the spectra measured from these detectors is a measurement of the antenna bandpass as there are no other optical or microstrip elements with a relevant frequency response. From the three 220 GHz modules there are seven out of twelve loss test detectors that yielded FTS measurements serviceable for this analysis, the reported antenna spectrum is the mean of these peak-normalized detector spectra. The spectrum we will compare the FTS results of the standard detectors with is product of the simulated filter and the peak-normalized measured antenna. These three spectra are shown together in Fig. \ref{filterxantenna}.

When looking at FTS spectra, we employ a different definition of band center and bandwidth than in Section \ref{Optimization}. In general, FTS spectra will be less smooth and have more features than model spectra so we use a definition that accounts for these non-idealities. We employ a power weighted mean definition for band center and a noise equivalent square band (NESB) definition of bandwidth.
\begin{align}
\label{BC}
    \langle\nu\rangle \equiv \frac{\int{S(\nu)\nu{d}\nu}}{\int{S(\nu){d}\nu}}
\end{align}
\begin{align}
\label{ESB}
    {\Delta\nu}\equiv\frac{\left(\int{S(\nu){d}\nu}\right)^2}{\int{S(\nu)^2{d}\nu}}
\end{align}
with $S(\nu)$ being the power response of the filter. The NESB bandwidth is defined as the bandwidth of a square filter that would have an equivalent signal-to-noise ratio as the given filter. These definitions are identical to those in Equation \ref{NET_defs} for calculating NET, the reason we describe the filters in section \ref{Optimization} with the simple definitions is because Equations \ref{BC} and \ref{ESB} are not invertible, i.e. we cannot plug in a band center and bandwidth and get out a filter. These definitions are also consistent with previous BICEP/\textit{Keck} work\cite{BICEPII}. 

\begin{figure}[]
\centering
\includegraphics[scale=0.5]{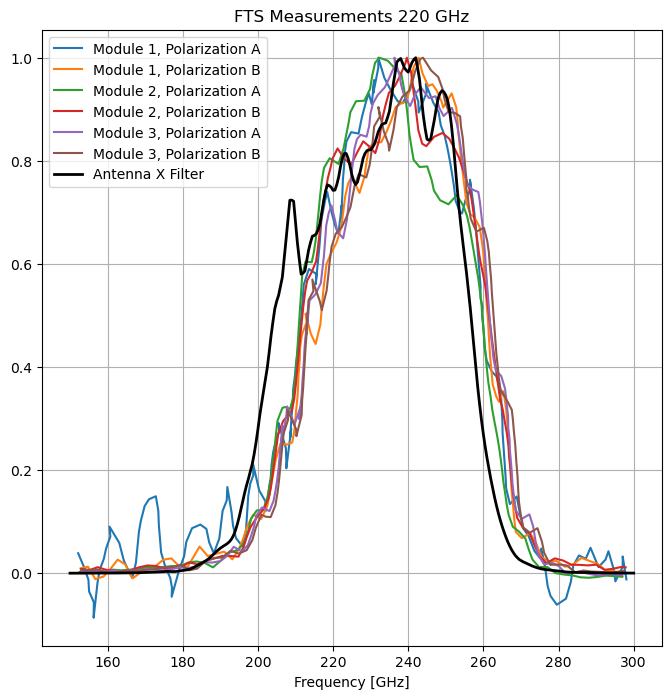}
\caption{The product of the measured antenna and simulated filter spectra (black) and peak-normalized mean passbands of both polarizations for the three deployed modules (colors).}
\label{FTS 220}
\end{figure}

The power-weighted band center of the 220 GHz antenna and filter product is 231.2 GHz and the NESB fractional bandwidth is 0.273. The mean passbands of all the working detectors from FTS measurements are shown along with the peak-normalized antenna-filter product in Fig. \ref{FTS 220}. There are six mean passbands here corresponding to three modules and two polarizations. The passbands are also peak-normalized.

Across the six populations shown in Fig. \ref{FTS 220}, we measured a mean power-weighted band center of 235.6 GHz with a mean standard deviation of 4.1 GHz, and a mean NESB fractional bandwidth of 0.252 with a mean standard deviation of 0.020, putting both of these values just over one standard deviation from the expected values. The FTS measurement yields for the six populations range from \(\sim \)54\% to \(\sim \)81\%, so all populations have a representative sample. We do not note any physical gradients in band center or bandwidth across any of the modules. 

One possible explanation for the discrepancy between the predicted and measured responses is an inaccurate relative permittivity or surface inductance parameter in the \textit{Sonnet} simulations. Another possibility is an inaccurate antenna measurement given the low sample size of the loss test pixels. It is important to note that this discrepancy has no significant impact on the ability of the 220 and 270 GHz BA receiver to achieve its science goals.

\begin{figure}[]
\centering
\includegraphics[scale=0.5]{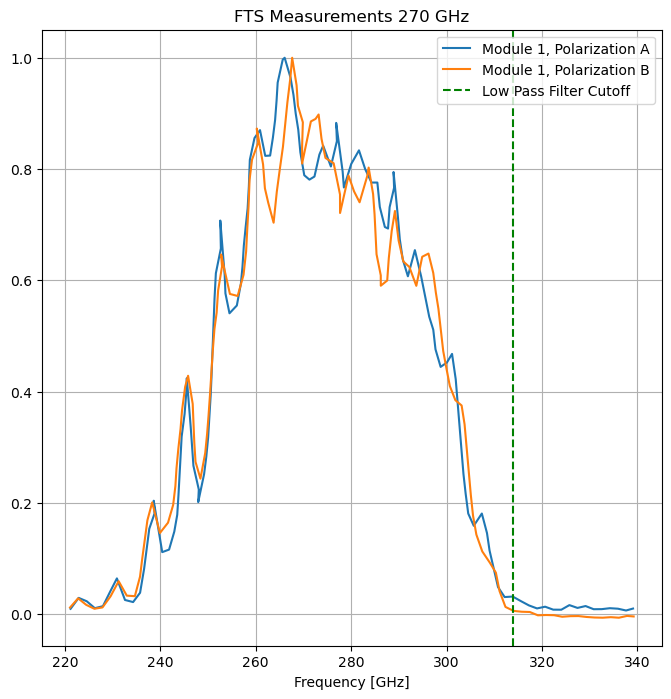}
\caption{The peak-normalized mean passbands of both polarizations for the one module 270 GHz module measured in lab.}
\label{FTS 270}
\end{figure}

As of September 2025, there have been no 270 GHz modules deployed to the South Pole. One 270 GHz detector module has been measured in lab but no loss test detector spectra were measured, so we are unable to make an accurate antenna measurement. The mean passbands are shown in Fig. \ref{FTS 270}. There is an optical metal-mesh low pass filter in the receiver to prevent blue leaks which is relevant for the 270 GHz spectra and also denoted in Fig. \ref{FTS 270}. This filter has a half power cutoff frequency of 314 GHz.

For the 270 GHz module, we measured a mean power-weighted band center of 274.0 GHz with a mean standard deviation of 3.05 GHz, and a mean NESB fractional bandwidth of 0.207 with a mean standard deviation of 0.022. We do not have high quality loss test data currently so we are unable to compare the measurements to expectation as with the 270 GHz measurements. The FTS measurement yield on the 270 GHz is lower at \(\sim \)35\% due to yield issues with the 270 GHz fabrication unrelated to the filters. We again do not note any physical gradients in band center or bandwidth on the module. Similar to the 220 GHz FTS results, these band passes are adequate for the science goals of the receiver.

 \section{Conclusion}
The BPFs used on the 220 GHz and 270 GHz BA receiver use a 3-pole pi-network design that can be fabricated with just two Nb films and without shunt inductors. The filters are designed and simulated with \textit{Sonnet}, and the band parameters are selected with a noise optimization considering loading from the atmosphere and the CMB. Three 220 GHz detector modules have been tested at the South Pole and one 270 GHz has been tested in lab. The 220 GHz expected spectrum, which is the product of the simulated filter and the antenna measurement from loss test pixels, is just outside one standard deviation from the FTS measurements in terms of both power-weighted band center and NESB fractional bandwidth. This discrepancy is potentially due to an inaccurate \textit{Sonnet} simulation parameter or an inaccurate antenna measurement. 
A similar comparison cannot currently be made for the 270 GHz spectra currently given the lack of an antenna measurement. Both the 220 and 270 GHz FTS measurements show spectra that will allow the 220 and 270 GHz receiver to achieve its science goals. 
Future work will include comparing the 270 GHz FTS measurements to the expected spectrum, accounting for the antenna bandpass as with 220 GHz.
\section*{Acknowledgments}
The BICEP/Keck experiments have been funded through
U.S. National Science Foundation grants most recently in-
cluding 2220444-2220448, 2216223, 1836010, and 1726917.
The research was carried out at the Jet Propulsion Laboratory, California Institute of Technology, under a contract with the National Aeronautics and Space Administration (80NM0018D0004). Focal plane development and
testing were supported by the Gordon and Betty Moore
Foundation at the California Institute of Technology. Readout
electronics were supported by the Canada Foundation for
Innovation grant to the University of British Columbia. The
computations in this paper were run on the Cannon cluster
supported by the FAS Science Division Research Computing
Group at Harvard University. The analysis effort at Stanford
University and the SLAC National Accelerator Laboratory was
partially supported by the Department of Energy. We thank
the staff of the U.S. Antarctic Program and in particular the
South Pole Station without whose help this research would not
have been possible. We also thank our winter-over operators:
Manwei Chan, Karsten Look, Calvin Tsai, Paula Crock, Ta
Lee Shue, Grantland Hall, Hans Boenish, Robert Schwarz,
Sam Harrison, Anthony DeCicco, Thomas Leps, Brandon
Amat, Nathan Precup, Steffen Richter, Thibault Romand, Danielle Simmons, Markus Ayasse, and Steven Jungst.


\vfill

\end{document}